\documentclass[aps,pre,showpacs,twoside,twocolumn,floatfix]{revtex4}

\usepackage{epsfig}
\usepackage{psfrag}
\usepackage{amsmath}
\usepackage{amssymb}
\usepackage{bm}
\usepackage{times}

\def\picdirectory{.}

\begin{document}
\title{Local equation of state and velocity distributions of a driven
  granular gas} 

\author{Olaf Herbst, Peter M\"uller, Matthias Otto, and Annette
  Zippelius}

\affiliation{Institut f\"ur Theoretische Physik, 
  Georg-August-Universit\"at, D--37077 G\"ottingen, Germany} 

\date{April 9, 2004}

\begin{abstract}
  We present event-driven simulations of a granular gas of inelastic
  hard disks with incomplete normal restitution in two dimensions
  between vibrating walls (without gravity).  We measure hydrodynamic
  quantities such as the stress tensor, density and temperature
  profiles, as well as velocity distributions.  Relating the local
  pressure to the local temperature and local density, we construct a
  local constitutive equation. For strong inelasticities the local
  constitutive relation depends on global system parameters, like the
  volume fraction and the aspect ratio. For moderate inelasticities
  the constitutive relation is approximately independent of the system
  parameters and can hence be regarded as a local equation of state,
  even though the system is highly inhomogeneous with heterogeneous
  temperature and density profiles arising as a consequence of the
  energy injection.  Concerning the local velocity distributions we
  find that they do not scale with the square root of the local
  granular temperature.  Moreover the high-velocity tails are
  different for the distribution of the $x$- and the $y$-component of
  the velocity, and even depend on the position in the sample, the
  global volume fraction, and the coefficient of restitution.
\end{abstract}


\maketitle


\section{Introduction}


The physics of vibro-fluidized granular materials is far from being
fully understood. In particular, the applicability of hydrodynamics
\cite{goldshtein95, grossman97, brey98, sela98, goldhirsch03} is still
an object of debate \cite{du95, tan98, kadanoff99}. Despite the
similarities to the hydrodynamics of elastic hard-sphere systems,
concerning e.g.\ the appearance of instabilities \cite{jaeger96b,
  livne02, livne02b, meerson02}, a main difference to ordinary fluids
is the fact that continuous energy injection is vital to maintain a
stationary state. Otherwise a gas of inelastic spheres would collapse,
even in the absence of gravity. The non-trivial nature of this
stationary state has been elucidated in experiments of vibrating
grains. Among the most striking features found are non-Gaussian
velocity distributions \cite{olafsen98, urbach99, losert99, rouyer00}
and cluster formation \cite{kudrolli97b, falcon99, falcon99b}.
Furthermore, several attempts have been made to extract either an
equation of state or more generally a scaling relation for the
thermodynamic variables \cite{warr95, falcon00}.

The specification of the driving mechanism is a crucial ingredient for
any model describing vibrated granular fluids. A rather simple
approach, though less appealing from an experimental point of view,
utilizes stochastic bulk heating by uncorrelated random forces, which
act on every particle at every instant of time \cite{williams96b,
  noije98d, noije99, barrat02c}.  In \cite{barrat01} random
restitution coefficients were considered with a probability
distribution allowing for values both smaller and greater than one.
However, this yields non-universal properties depending on the
specific form of this distribution. The multiplicative bulk driving
defined in terms of stochastic collision rules in \cite{cafiero00} is
similar in spirit. As compared to bulk driving, a driving mechanism
which acts only at a boundary of the system is much closer to
experiments. In \cite{grossman97} the energy influx at the boundary
has been modeled by a heuristically motivated ansatz for the heat
current, while \cite{brey98c,risso02} assume heating at the boundary
through thermal walls.  In the present paper the driving mechanism
consists in incrementing a particle's velocity by a constant amount
after every collision with a wall. This way of driving was used before
in e.g.\ \cite{brey00c, barrat02b}, and with slight modifications in
\cite{drossel98, kumaran98, kumaran99}.

One theoretical approach to study granular systems at fluid densities
rests on kinetic theory, see e.g.\ \cite{noije98, brey99, dufty01,
  goldhirsch03} and references therein. Particular applications to
driven granular gases can be found in \cite{brey98c, noije98d,
  cafiero00, krapivsky01, barrat01}.  Much of this work has been based
on the Boltzmann equation, modified for inelastic collisions. One
method to solve this nonlinear equation is based on the local
equilibrium distribution \cite{jenkins83}, which is only known for the
elastic case.  For systems with strongly inelastic collisions, the
stationary state is unknown so that a systematic discussion of
transport properties within kinetic theory is severely hampered.

Hydrodynamic studies \cite{grossman97, brey00c, livne02b, meerson02}
have been motivated partly by the search for an understanding of
temperature and density profiles \cite{grossman97}, but also by
experiments on hydrodynamic-like instabilities \cite{jaeger96b}. In
the hydrodynamic approach the question arises, how to relate the
pressure to the density and temperature. Several equations of state
for the global quantities have been proposed, either interpolating
between the high and low density limit \cite{grossman97} or invoking
the Boltzmann equation \cite{barrat02b} for inelastically colliding
particles.  It is not clear, whether an equation of state also holds
for the local hydrodynamic fields in a strongly driven,
non-equilibrium system, where the fields are strongly inhomogeneous.

Most simulations of vibrated granular gases have been based either on
event-driven molecular dynamics \cite{luding94c, mcnamara97,
  barrat02b} or the direct-simulation Monte Carlo method
\cite{brey98c, brey00c, puglisi99, baldassarri01b}.  If the system is
driven through the boundaries, inhomogeneous density and temperature
profiles are measured. For low densities the computed temperature
profiles agree well with hydrodynamic theory \cite{grossman97,
  brey00c}, whereas for moderate or high densities the profiles are
not well understood with the exception of almost elastically colliding
particles. The full stress tensor, including potential contributions,
has only been computed for freely cooling systems \cite{luding00b}. In
simulations of driven granular gases \cite{brey98c} the collisional
part of the stress tensor has not been measured, and in
\cite{barrat02b} the stress tensor is not measured directly but
instead computed from a local equation of state.

Inspired by experiment \cite{rouyer00}, a lot of emphasis has been put
on the tails of the velocity distribution functions, which were found
to be overpopulated as compared to a Maxwellian. In fact, all
intermediate types of decay between a Gaussian and an exponential were
observed \cite{puglisi99, barrat02b, zon03, brey03a}.  In addition
mixtures \cite{barrat02b, barrat02, paolotti02} and rough spheres
\cite{luding95b} have been investigated as well as hydrodynamic
instabilities such as convection \cite{luding94b,ramirez00} and
pattern formation \cite{luding96e}.

In this paper, we present results from event-driven simulations of
inelastic spheres in two dimensions confined between two vibrating
walls without gravity. Our focus is on the stationary state, which is
reached when dissipation by particle collision equals energy injection
due to the vibrating walls. In the stationary state, density and
temperature profiles are shown to be strongly inhomogeneous due to the
driving walls---even in a range of parameters, where clustering is
only a minor effect. This has led us to

(a) derive a constitutive equation by relating the measured
hydrodynamic fields, granular temperature $T(x)$, volume fraction
$\phi(x)$ and pressure $p(x)$, at each point $x$ and

(b) check whether this constitutive equation is universal or depends
on the global system parameters of the model, like the aspect ratio of
the cell, the overall volume fraction or the coefficient of
restitution of the disks. For moderately inelastic systems ($\alpha =
0.9$) the constitutive equation is (almost) independent of the
remaining global system parameters so that the constitutive equation
can be interpreted as a local equation of state for the a driven
granular gas in the stationary state, even though the latter is highly
inhomogeneous with heterogeneous temperature and density profiles.  In
contrast, for strongly inelastic systems ($\alpha =0.5$) the
constitutive equation depends significantly on the global system
parameters so that the concept of a local equation of state cannot be
sustained in this case.

We furthermore discuss the one-particle distribution function in the
stationary state and show that 

(c) the local distribution $f_x(x,v_x)$ of $v_x$, the velocity in the
direction of driving, is not a function of the rescaled variable
$v_x/\sqrt{T_x(x)}$ alone. Similarly, curves of $f_y(x,v_y)$ cannot be mapped
onto a master curve for different $x$, when plotted against
$v_y/\sqrt{T_y(x)}$.  Here $T_i(x)$ denotes the local granular temperature
associated with the translational motion in the $i$-direction. We find
deviations from scaling at small and large arguments.

(d) The local velocity distributions $f_{x}$ and $f_{y}$ have high-velocity
tails whose decay ranges from stretched exponential to almost Gaussian. The
particular type of decay depends on the position in the sample, the overall
particle density and the coefficient of restitution. Furthermore, the
decay of $f_{x}$  for large velocities is generally different
from that of $f_{y}$.

Taken together, one has to acknowledge that the stationary state of a
driven granular gas is by no means universal. It shows peculiar
features that depend on the precise values of the system parameters
in contrast to the corresponding elastic system.

The outline of the paper is as follows.  In Sec.\ \ref{model} we
introduce the model, specify the driving and define the observables.
We briefly discuss balance of energy input through the walls and
energy dissipation in binary collisions in Sec.\ 
\ref{sec:DimAnalyseUndEnergiebilanz}. Subsequently we present data for
the profiles of the density, the temperature and the components of the
stress tensor (Sec.\ \ref{sec:hydr.fields}). In Sec.\ 
\ref{sec:loc.equ.state} we relate the local density, pressure and
temperature to derive ``experimentally'' an equation of state and
check for its universality.  Finally, in Sec.\ \ref{sec:vel.distr.}
we discuss velocity distribution functions and their scaling behavior
and present conclusions in Sec.\ \ref{sec:conclusions}


\section{Model and Observables}
\label{model}

\begin{figure}
  \begin{center}
    \leavevmode \psfig{file=\picdirectory/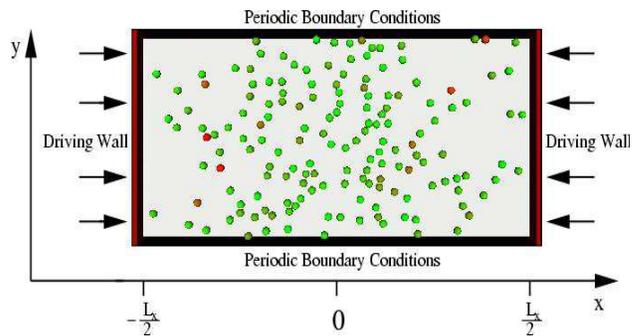, clip=,
      width=0.95\columnwidth, height=0.5\columnwidth}
    \caption{Model of N disks, driven in the $x$-direction with
      periodic boundary conditions in the $y$-direction.
      \label{fig:Schnappschuss}}
  \end{center}
\end{figure}

We investigate a driven granular gas in 2 dimensions consisting of $N$
identical inelastic smooth hard disks of diameter $a$ and mass $m$
which are confined to a rectangular box with edges of length $L_x$ and
$L_y$.  The gas is driven through the walls perpendicular to the
$x$-direction, which vibrate in an idealized saw-tooth manner (see
below), while periodic boundary conditions are imposed in the
$y$-direction.  Fig.~\ref{fig:Schnappschuss} shows a typical snapshot.
The gas evolves in time through ballistic
center-of-mass motion, binary inelastic collisions and
particle-wall collisions.

\subsection{Binary collisions}


The inelastic nature of inter-particle collisions is the most
important characteristic of granular media.  As is often done we
assume that it can be taken into account by a constant coefficient of
normal restitution and briefly recall the collision rules for binary
collisions, see e.g. \cite{goldhirsch93, brilliantov96}.

The unit-vector from the center of sphere two to the center of sphere
one is denoted by $\hat{\bm{n}} := (\bm{r}_{1}-\bm{r}_{2})/|\bm{r}_{1}
-\bm{r}_{2}|$ where $\bm{r}_{i}$ is the position vector of the center
of mass of particle $i$.  The center-of-mass velocities before a
collision are denoted by $\bm{v}_1$ and $\bm{v}_2$, and the relative
velocity by $\bm{v}_{12}=\bm{v}_1- \bm{v}_2$.  Post-collisional
quantities are primed. The relative velocity after a collision is
assumed to obey
\begin{equation}
  \label{eq:rauku4}
  \hat{\bm{n}}\cdot\bm{v}'_{12} =  -\alpha  ~\hat{\bm{n}}\cdot\bm{v}_{12}
\end{equation}
where $\alpha \in [0,1]$ is the (constant) coefficient of normal
restitution.  The value $\alpha=1$ describes elastic collisions with
energy conservation, while for $\alpha < 1$ energy will not be
conserved but decreased in each collision.
 
The constitutive equation (\ref{eq:rauku4}) plus momentum conservation
determines the post-collisional velocities for the disks
\begin{equation}
  \label{eq:freddy} 
  \bm{v}_1' = \bm{v}_1+ \Delta \bm{v}_{\rm pp},  \qquad \bm{v}_2' =
  \bm{v}_2- \Delta \bm{v}_{\rm pp}
\end{equation}
where
\begin{equation}
  \label{eq:Deltas}
  \Delta \bm{v}_{\rm pp} =  - \frac{(1+\alpha)}{2}
  (\hat{\bm{n}}\cdot\bm{v}_{12})~ \hat{\bm{n}}.
\end{equation}
The model could easily be extended to rough spheres by also specifying
the disks' moment of inertia as well as a coefficient of tangential
restitution $\beta_0$ and Coulomb friction $\mu$ \cite{walton86,
  herbst00}, but here we restrict ourselves to smooth spheres so that
rotational motion is decoupled from translational motion.


\subsection{Driving} 


When a particle collides with a driving wall, energy is injected into
the system.  This can be modeled in different ways, for example by
drawing a new velocity from a Maxwellian distribution of a given wall
temperature \cite{grossman97} or by assuming that the wall has a
coefficient of normal restitution that is greater than one. Both of
these mechanisms have no close experimental equivalents, though.  A
more realistic model is to assume a vibrating wall moving either in a
symmetric (e.g. sinusoidal) or in an asymmetric (e.g.  saw-tooth) way.
In addition, this can be combined with a normal (and also tangential)
coefficient of restitution \cite{mcnamara97, kumaran98, kumaran99b}.
In this article we refrain from the latter and restrict ourselves to
saw-tooth driving of the walls in the limit of vanishing amplitudes
$A$ and diverging frequency $\nu$ such that $A\nu=: v_{\rm drive}/2$
is a constant.  This ensures that the driving walls are always located
at the same positions and leads to the following simple expression for
a particle's change of velocity due to a collision with the left/right
wall
\begin{equation} 
  \label{eq:freddyandDeltasWall} 
  \bm{v}' = \bm{v}+ \Delta \bm{v}_{\rm pw} \quad 
  \mbox{where} \quad \Delta \bm{v}_{\rm pw} =
  (- 2 v_x \pm v_{\rm drive}) \bm{e}_{x}
\end{equation}
and $\bm{e}_{x}$ stands for the unit vector in $x$-direction.


\subsection{Observables from simulations}
\label{sec:observables}


Following e.g. \cite{lubachevsky91, luding98f} we performed event-driven
simulations in two dimensions with periodic boundary conditions in the
$y$-direction and two identical idealized vibrating elastic walls in the
$x$-direction.  We initialize the system by placing the disks on a triangular
lattice with a Gaussian velocity distribution.  To let the correlations of the
initial state relax, the system is evolved elastically with periodic boundary
conditions in the $y$-direction and elastic non-vibrating walls in the
$x$-direction for an average of 100 collisions per particle.  Then we switch
on the driving of the left and right wall and the dissipation for
particle-particle-collisions.  Before we start measuring the observables we
let the system relax further until, at time $t_0$, it has reached a stationary
state, as indicated by the total kinetic energy, which fluctuates
around a time-independent mean value.

To measure hydrodynamic fields in the stationary state we subdivide
our box into cells $\mathcal{V}_{\bm{r}}$ of area
$|\mathcal{V}_{\bm{r}}|$, centered at position $\bm{r}=(x,y)$. We
then count the number of particles $f(\bm{r}, v_x,v_y,t)
|\mathcal{V}_{\bm{r}}| dv_x dv_y$ at time $t$ in cell
$\mathcal{V}_{\bm{r}}$ with an $x$-component of the velocity between
$v_x$ and $v_x+dv_x$ and $y$-component of the velocity between $v_y$
and $v_y+dv_y$.  Such a local observable fluctuates as a function of
time.  To eliminate these fluctuations we average over a (long) time
interval of length $\tau$ and compute
\begin{equation}
  f_{\rm stat}(\bm{r}, v_x,v_y) 
  := \frac{1}{\tau}\int_{t_0}^{t_0+\tau} dt~ f(\bm{r}, v_x,v_y,t) \,.
\end{equation}
Of particular interest is the particle density
\begin{equation}
  \label{density}
  \rho(\bm{r}) := \int_{\mathbb{R}} dv_x  \int_{\mathbb{R}} dv_y \;
  f_{\rm stat}(\bm{r}, v_x,v_y) 
\end{equation}
or the area fraction $\phi(\bm{r}) := \rho(\bm{r})\pi a^2/4$ and the
two components $i=x,y$ of the granular temperature
\begin{equation} 
  T_i(\bm{r}) := \frac{m}{\rho(\bm{r})}\int_{\mathbb{R}} dv_x 
  \int_{\mathbb{R}} dv_y\; f_{\rm stat}(\bm{r}, v_x,v_y)\;
  (v_i-V_i(\bm{r}))^2 
\end{equation}
where
\begin{equation}
  \bm{V}(\bm{r}) :=
  \frac{1}{\rho(\bm{r})}\int_{\mathbb{R}} dv_x  
  \int_{\mathbb{R}} dv_y\; f_{\rm stat}(\bm{r}, v_x,v_y)\,  \bm{v}
\end{equation}
is the velocity field. The total granular temperature in 2
dimensions is defined as $T(\bm{r}) := (T_x(\bm{r})+T_y(\bm{r}))/2$.

The stress tensor at position $\bm{r}$ and time $t$ has a kinetic
contribution and one that is due to the interactions between the
particles $\boldsymbol{\sigma}(\bm{r},t) := \boldsymbol{\sigma}^{\rm
  kin}(\bm{r},t)+ \boldsymbol{\sigma}^{\rm int} (\bm{r},t)$. The
kinetic part is given by
\begin{align}
  \sigma_{ij}^{\rm kin}(\bm{r}) := -m\int_{\mathbb{R}} dv_x &
  \int_{\mathbb{R}} dv_y\; f_{\rm stat}(\bm{r}, v_x,v_y) \nonumber\\
  & \times [v_i - V_{i}(\bm{r})] [v_j -V_{j}(\bm{r})] \,.
\end{align}
If the particles interact through (finite) forces the contribution due
to interactions is given by the correlation of the particles' relative
positions and the forces between them. For hard-core interactions
there are no forces, and one has to consider the momentum transfer in
a small time interval instead \cite{luding98f, glasser01}:
Suppose there is a collision at time $t$ of particle $k$ with another
particle, then the momentum change of particle $k$ will contribute to
the stress tensor an amount proportional to $l_i^k(t) \Delta
p_j^k(t)$.  Here $l_i^k(t)$ is the $i$-th component of the vector of
length $a/2$ pointing from the center of disk $k$ to its collision
contact point, and $\Delta p_j^k(t)$ is the $j$-th component of the
momentum change of particle $k$ during this collision.
To compute the collisional part of the stress tensor we need the total
change of momentum in the time interval $[t-\Delta t,t]$ in the cell
$\mathcal{V}_{\bm{r}}$ so that we have to keep track of all collisions
$n$ occurring at times $t_n \in [t -\Delta t, t]$, for which at least
one collision partner $k_n$ (i.e. its center of mass) is located in
cell $\mathcal{V}_{\bm{r}}$ at time $t_n$ \cite{luding98f}
\begin{equation}
  \label{eq:def.stress.tensor}
  \sigma_{ij}^{\rm int}(\bm{r},t)=
  \frac{1}{\Delta t} \frac{1}{|\mathcal{V}_{\bm{r}}|}
  \sum_{t_n}\sum_{k_n}  l_i^{k_n}(t_n) \Delta p_j^{k_n}(t_n)\,.
\end{equation}
The particle number $k_n$ of each such collision can take on one or
two values, depending on whether one or both collision partners are
located in cell $\mathcal{V}_{\bm{r}}$.  Similar to the other
hydrodynamic fields, in a second step $\boldsymbol{\sigma}^{\rm
  int}(\bm{r},t)$ is averaged over time in order to get the
collisional part of the stress tensor in the stationary state
\begin{equation}
  \label{longtimestress}
  \sigma_{ij}^{\rm int}(\bm{r}) =  \frac{1}{\tau}
  \int_{t_0}^{t_0+\tau}dt\; \sigma_{ij}^{\rm int}(\bm{r},t) \,.
\end{equation}
The corresponding local pressure
$p(\bm{r}):=-\mbox{tr}\left(\boldsymbol{\sigma}(\bm{r})\right)/2$ 
is defined as usual as the negative trace of the stress tensor divided
by the space dimension.

Besides hydrodynamic fields we have also measured velocity
distributions in Sec.~\ref{sec:vel.distr.}. They are readily obtained
by integrating out the relevant variables in the stationary-state
distribution function $f_{\mathrm{stat}}$. A different method, which
is better suited to determine high-velocity tails, for example, will
be presented in Sec.~\ref{sec:vel.distr.} below.

Coarse-grained measurements in space and time of certain observables
may depend on the coarse-graining resolution. For example,
this was demonstrated for the stress tensor in shear-flow driven
granular systems in \cite{goldhirsch98, glasser01}. Our measurements
of observables in the stationary state, however, should not suffer
from such effects for two reasons: First, we could not detect any
significant non-zero local velocity field in the
simulated systems and, second, because of the long-time average needed
to obtain stationary-state quantities. 

Moreover, in the simulations we have never found any significant
dependence of the (long-time-averaged) hydrodynamic fields on $y$, the
coordinate parallel to the driving walls. Yet, stripe states, which are
homogeneous in $y$, but have an enhanced density in the middle of the
sample, are known \cite{livne02, livne02b} to exhibit instabilities
with respect to density fluctuations in $y$. A marginal stability
analysis \cite{livne02b} of granular hydrodynamics (for $\alpha$ close
to one) determines the conditions under which such phenomena occur.
As far as a comparison can be made, our systems fall into the stable
region. Hence, we choose
the cells $\mathcal{V}_{\bm{r}}$ as stripes along the $y$-direction,
for which we write $\mathcal{V}_{x}$, and compute the hydrodynamic
fields with spatial resolution in $x$-direction only. Thus, we also
write $\rho(x)$ instead of $\rho(\bm{r})$ and change the notation
accordingly for all other quantities. In the simulations the stripes
$\mathcal{V}_{x}$ were all chosen to have equal width $L_{x}/201$, the
temporal resolution of our measurements was set to $\Delta t=1$, and
the long-time average involves typically $10^7$ -- $10^9$ collision
events.

Another instability of stripe states which is related to oscillations of the
central dense cluster in $x$-direction \cite{khain04} will be shortly
addressed at the end of Subsec.~\ref{subsec:DensityTempProfiles}.


\section{Dimensional analysis and energy balance}
\label{sec:DimAnalyseUndEnergiebilanz}
  

The model system contains three independent length scales: the
diameter $a$ of a disk and the box sizes $L_x$ and $L_y$. In addition,
there is one independent velocity scale, the driving velocity $v_{\rm
  drive}$, and one independent mass scale, the mass of a disk $m$.
Together with the initial positions and velocities of the particles
that exhausts all dimensional quantities entering the time evolution
of the system.
  
We would like to describe the system using dimensionless variables
that do not depend on the initial conditions since we expect
stationary states to be independent thereof.  Therefore, we will
measure all lengths in units of the particle diameter $a$, all times
in units of $a/v_{\rm drive}$ and all energies in units of $mv_{\rm
  drive}^2$. Note that there are no other time and energy
scales. Thus, we introduce dimensionless variables: box 
sizes $\tilde{L}_x=L_x/a $ and $\tilde{L}_y=L_y/a $, granular
temperatures $\tilde{T}_x=T_x/(mv_{\rm drive}^2)$ and
$\tilde{T}_y=T_y/(mv_{\rm drive}^2)$, and the stress tensor
$\tilde{\boldsymbol{\sigma}} = \boldsymbol{\sigma}\, a^2 / (m v_{\rm
  drive}^2)$.
In the stationary state all dimensionless variables like $\tilde{T}$
and $\tilde{\boldsymbol{\sigma}}$, are \emph{independent of the
driving velocity} and only depend on position $\bm{r}$ and the
remaining 4 independent dimensionless system parameters, which
characterize the system completely: the number of disks $N$, the two
edges of length of the system $\tilde{L}_x$ and $\tilde{L}_y$ in units of $a$,
and the coefficient of normal restitution $\alpha$, which is a
dimensionless material constant. 

For simplicity of the notation we refrain from indicating
dimensionless quantities by a tilde from now on. This should not cause
confusion, because quantities having a physical dimension will not
occur any more in the rest of the paper (except for the appendix).

We are interested in the macroscopic limit, which
is taken such that $N \to \infty$ and $L_y \to \infty$ with a fixed
line density $\lambda := N/L_y$ of particles. Thus, in the
macroscopic limit the number of system parameters is further reduced
to $L_x$, $\lambda$, and $\alpha$.
It is important to keep $L_x$ finite, 
otherwise energy balance would not work: energy input occurs only at
boundaries, while energy dissipation is a bulk phenomenon.

\begin{figure}
  \begin{center}
    \leavevmode
    \psfrag{Tq}{$\overline{T}$}
    \psfrag{psi1}{$\psi^{-1}$}
    \psfrag{Theory-simple energy balance}{\tiny Simple energy balance
      (\ref{eq:EnBilanzGlSimple})} 
    \psfrag{Theory-refined version}{\tiny Refined energy balance
      (\ref{eq:EnBilanzGlFullPseudoLiouville})} 
    \epsfig{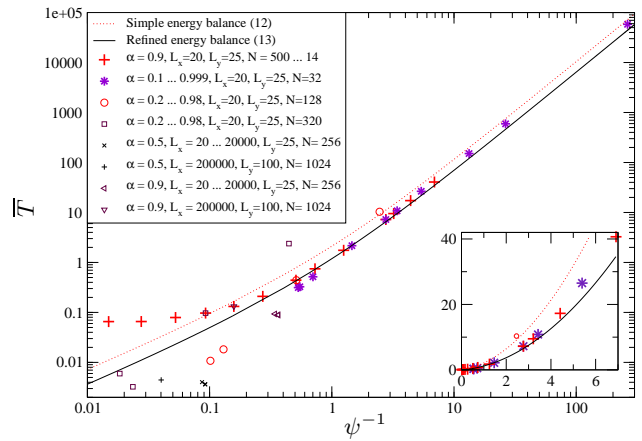}
    \caption{Global granular temperature $\overline{T}$ as a function of the
      parameter $\psi^{-1}$ defined below (\ref{eq:EnBilanzGlSimple}).
      Comparison between simulations and the simple energy-balance
      argument, Eq.~(\ref{eq:EnBilanzGlSimple}), as well as with the
      refined version, Eq.~(\ref{eq:EnBilanzGlFullPseudoLiouville}).
      The inset shows the same graph but on on a non-logarithmic
      scale.
      \label{fig:TvonPsi}}
  \end{center}
\end{figure}
 
It is instructive to estimate the average or global granular
temperature $\overline{T}:= \int_{-L_{x}/2}^{L_{x}/2} dx\, \rho(x)\,
T(x)$ by balancing the energy input at the walls and the energy loss
due to particle collisions in the bulk \cite{kumaran98, mcnamara98b}.
As shown in the appendix, we find that the granular temperature in the
stationary state is independent of the initial data and given by
\begin{equation} 
  \label{eq:EnBilanzGlSimple}
  \overline{T}= \biggl(\frac{2}{\pi}\biggr)^3 \psi^{-2} \, 
  \left( 1 + \sqrt{1+ (\pi/2)^{2} \psi} \, \right)^2 
\end{equation}
where $\psi := \sqrt{2} \chi \lambda (1-\alpha^2)$ and $\chi$ stands
for the pair correlation at contact of the corresponding elastic
system, which we estimate by the Henderson approximation
(\ref{chiapprox}). In Fig.~\ref{fig:TvonPsi} we plot the global
granular temperature $\overline{T}$ as a function of $\psi^{-1}$ and
compare it with simulations. For small values of $\psi^{-1}$ the
simulation deviates significantly from this simple theory's
prediction. For large values of $\psi^{-1}$ (dilute and/or
quasi-elastic systems) the agreement is reasonable. In addition, we
show the result
\begin{equation}
  \label{eq:EnBilanzGlFullPseudoLiouville}
  \overline{T} =  \frac{1}{2\pi\psi^2} \, \left( 1+ \sqrt{1+ \psi/2}\,
  \right)^2 
\end{equation}
of a more refined calculation, which uses the
pseudo-Liouville-operator approach to kinetic theory and will be
presented elsewhere. Eq.\ (\ref{eq:EnBilanzGlFullPseudoLiouville})
yields a better agreement with the data for intermediate values of
$\psi^{-1}$.

It is worth noting that the same type of argument which led to
(\ref{eq:EnBilanzGlSimple}) also predicts that there is no generic
stationary state for systems with the multiplicative driving mechanism
(\ref{general.wall.coll}), described by $v_{\rm drive}=0$ and a
coefficient of restitution bigger than one for particle-wall
collisions. Instead, such systems either cool down or heat up
according to Haff's law \cite{haff83}, see
Eq.~(\ref{eq:appendix.essentiallyHaff}) in the appendix. This was also
confirmed by simulations (not presented).


\section{Hydrodynamic Fields in Simulations}\label{sec:hydr.fields}


In this section we discuss the results of our simulations for the
hydrodynamic fields, as computed from Eqs.~(\ref{density}) --
(\ref{longtimestress}). Due to the absence of a local velocity field
$\bm{V}$, these equations simplify accordingly. We have
performed simulations for a wide range of system parameters $\alpha$,
$\phi_{0}$, $L_{x}$, $L_{y}$, and present examples thereof below. We
then go on in Sec.~\ref{sec:loc.equ.state} to discuss the question of
a local equation of state, relating $p(x)$, $\rho(x)$, and
$T(x)$.


\subsection{Density and temperature profiles}
\label{subsec:DensityTempProfiles}
  
\begin{figure}
  \begin{center}
    \leavevmode \epsfig{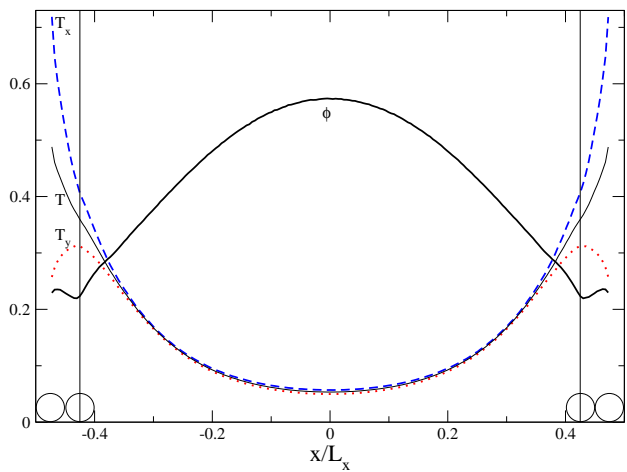}
    \caption{Spatial profiles of the
      area fraction $\phi$ and granular temperatures $T_x$, $T_y$, and
      $T$ from simulations [$\alpha =0.9$, $L_{x}=20$, $\lambda
      =10.24$, $N=256$, $\phi_{0} = 0.4$, corresponding to a mean-free
      path $\ell/L_{x} \approx 0.02$].
      \label{fig:dens.temp.kes.2}}
  \end{center}
\end{figure}

We first present data for the density and temperature to demonstrate
that the system is strongly inhomogeneous even for collisions with
$\alpha=0.9$ and moderate densities. The parameters have been chosen
as $L_x=20$, $L_y=25$, and $N=256$ so that the global area fraction is
$\phi_{0}:= \pi a^{2}N/(4V) = 0.4$. In Fig.~\ref{fig:dens.temp.kes.2}
we show the local area fraction $\phi(x)$, the $x$- and $y$-component
of the granular temperature, $T_x(x)$ and $T_y(x)$, as well as the
isotropic temperature $T(x)=[T_x(x)+T_y(x)]/2$. We note that all
quantities are symmetric in $x$ as expected. Except for a small area
of approximately one disk diameter next to the walls (indicated by the
vertical lines), the area fraction is a monotonic function for either
half of the system with the maximal value in the middle of the system.
The temperatures $T_x$, $T_y$, and $T$ are monotonic as well with the
lowest temperatures in the middle of the sample.

The reason for the increased density next to the wall is an effective
attractive potential of the wall due to entropic effects: Once a disks
gets closer to the wall than one disk diameter, it can only receive
hits from within the box but no hits from the direction of the wall.
Thus the particle is pushed closer to the wall \cite{rogers84,
  hansen86}. This effect is partially compensated by the driving
walls, which add momentum to any particle hitting the walls.
 
To support this explanation we have also investigated systems of half
the size $-0.5 \le x/L_x \le 0$, half the number of particles and with
an elastic wall at $x=0$.  The resulting hydrodynamic fields (not
shown) are almost identical to the ones in
Fig.~\ref{fig:dens.temp.kes.2} except for a small region of about one
diameter close to $x=0$ where the aforementioned effect is
particularly visible.

In systems exhibiting stripe states, we have sometimes observed an
oscillatory instability of a central dense cluster in the
$x$-direction \cite{khain04}, in particular if $\alpha$ is low and
$L_{x}$ is large. However, these oscillations occur on far shorter
time scales compared to the time interval $\tau$, over which we
average to obtain stationary-state quantities. Hence, these
oscillations are completely averaged out in the presented data.


\subsection{Stress tensor and pressure}

\begin{figure}
  \begin{center}
    \leavevmode \epsfig{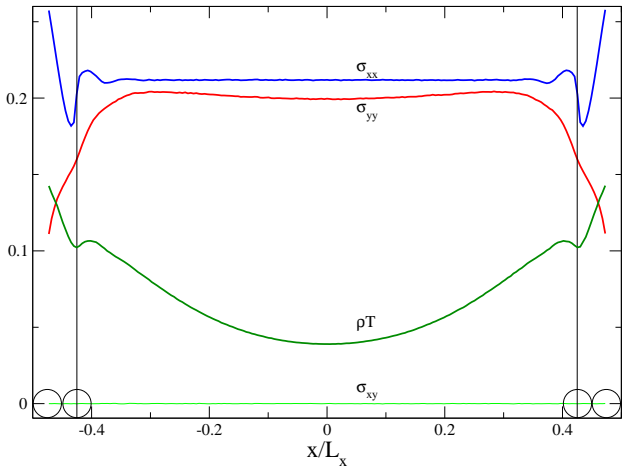}
    \caption{Spatial profiles of the stress tensor
      components $\sigma_{xx}$, $\sigma_{yy}$, $\sigma_{xy}$, and the
      product $\rho T$ for the system of Fig.~\ref{fig:dens.temp.kes.2}.
      For all $x$ away from the boundary layers, $\rho T$ is hardly to
      distinguish from $\rho T_{x}$ and $\rho T_{y}$ (not shown).
      \label{fig:stresses.kes.2}}
  \end{center}
\end{figure}
\begin{figure}
  \begin{center}
    \leavevmode
    \epsfig{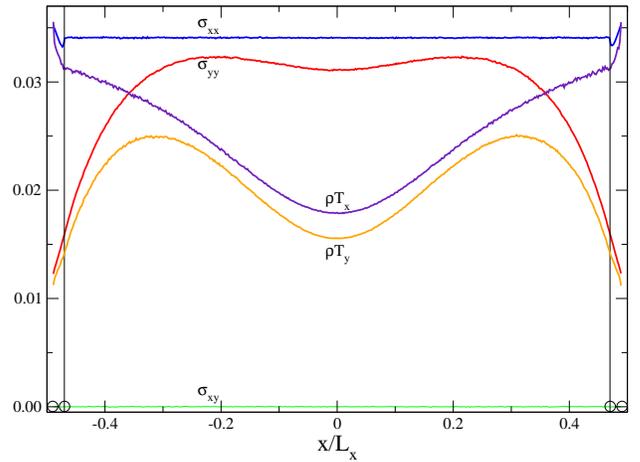}
    \caption{Same as Fig.~\ref{fig:stresses.kes.2}, but for a less
      dense system [$\alpha =0.9$, $L_{x}=50$, $\lambda
      =9.6$, $N=240$, $\phi_{0} = 0.15$, $\ell/L_{x} \approx 0.03$].
      \label{stresses.kim.4a}}
  \end{center}
\end{figure}
\begin{figure}
  \begin{center}
    \leavevmode
    \epsfig{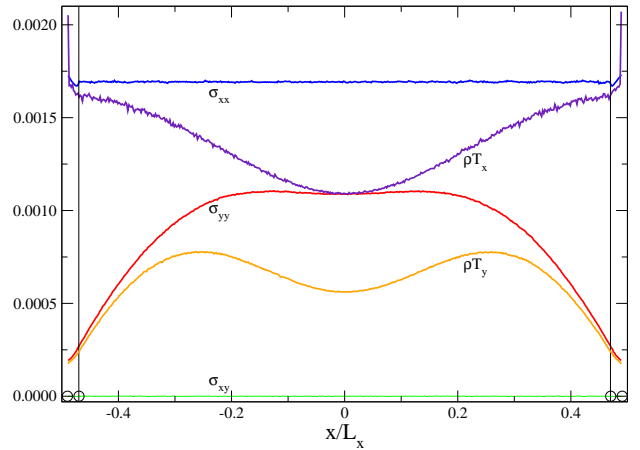}
    \caption{Same as Fig.~\ref{stresses.kim.4a}, but for $\alpha=0.5$.
      \label{stresses.kim.4}}
  \end{center}
\end{figure}

In Figs.~\ref{fig:stresses.kes.2} -- \ref{stresses.kim.4} we show the
components of the stress tensor $\sigma_{xx}(x)$, $\sigma_{yy}(x)$,
and $\sigma_{xy}(x)$ for two different sets of parameters.
Fig.~\ref{fig:stresses.kes.2} is typical for systems with quite
elastic collisions ($\alpha=0.9$) and moderate densities
($\phi_{0}=0.4$), whereas Figs.~\ref{stresses.kim.4a}
and~\ref{stresses.kim.4} show rather dilute systems ($\phi_{0}=0.15$)
at two different inelasticities, a moderate one ($\alpha=0.9$, as in
Fig.~\ref{fig:stresses.kes.2}) and a strong one ($\alpha=0.5$). The
off-diagonal component $\sigma_{xy}$ of the stress tensor is always
vanishingly small. The $xx$-component $\sigma_{xx}$ is constant within
the sample except for a boundary layer close to the driving walls,
whereas Fig.~\ref{fig:stresses.kes.2} displays a dip of $\sigma_{yy}$
in the center of the sample, which is more pronounced for the more
dilute system in Fig.~\ref{stresses.kim.4a} and hardly visible in the
strongly inelastic, dilute system of Fig.~\ref{stresses.kim.4}.
Furthermore, $\sigma_{yy}$ increases considerably over a broad range
in Fig.~\ref{stresses.kim.4}, when moving from a driving wall towards
the center of the system. Currently, the origin of both
$x$-dependences of $\sigma_{yy}$ is not clear to us.  While the
stationary-state condition
$\boldsymbol{\nabla}\cdot\boldsymbol{\sigma}=0$ and homogeneity in $y$
require $\sigma_{xx}$ and $\sigma_{xy}$ to be constant in $x$ on
general grounds, this is not the case for $\sigma_{yy}$. We have
carefully checked that the $x$-dependence of $\sigma_{yy}$ is not
caused by a shear instability associated with a non-zero velocity
field $\bm{V}(x)$ in the system. Thus, there are normal stresses
present in the simulated systems, and they depend on $x$.

If the equation of state of the ideal gas held, the local pressure
$p(x)$ would be related to the local temperature and density according
to $p(x)= \rho(x) T(x) = (4/\pi)\phi(x) T(x)$. For our system the
kinetic part of the stress tensor is diagonal and each component is
simply related to the corresponding temperature $\sigma^{\rm
  kin}_{xx}(x)=\rho(x) T_x(x)$ and $\sigma^{\rm kin}_{yy}(x)=\rho(x)
T_y(x)$.  Hence the difference between the measured stress tensor and
the ideal gas behavior is due to collisions.  From
Figs.~\ref{fig:stresses.kes.2} -- \ref{stresses.kim.4} it can be seen
that the stress tensor deviates strongly from the ideal gas behavior.
Consequently, collisions contribute significantly.  Here the
collisional part has been measured directly in the simulations and not
only estimated by approximate theories as has been done elsewhere
\cite{brey00c, barrat02b}.

We have also estimated the global mean free path $\ell$ for our
simulated systems according to Eq.~(9) in \cite{grossman97}, which
expresses $\ell$ solely as a function of the global volume fraction
$\phi_{0}$. For the denser system with $\phi_{0}=0.4$, shown in
Figs.~\ref{fig:dens.temp.kes.2} and~\ref{fig:stresses.kes.2}, this
gives $\ell\approx 0.5$ (in units of the diameter of the disks, that
is $\ell/L_{x} \approx 0.025$). For the thinner systems with
$\phi_{0}=0.15$, shown in Figs.~\ref{stresses.kim.4a}
and~\ref{stresses.kim.4}, one gets $\ell\approx 1.65$, that is
$\ell/L_{x} \approx 0.033$. Thus, in all cases the mean free path
$\ell$ is much smaller than the scales governing the spatial
variations of the hydrodynamic fields, which are of order $L_x$.

\subsection{Density scaling}
\label{sec:densprofile}

\begin{figure}
  \begin{center}
    \leavevmode 
    \psfrag{Grossman}{\footnotesize Ref.~\cite{grossman97}}
    \epsfig{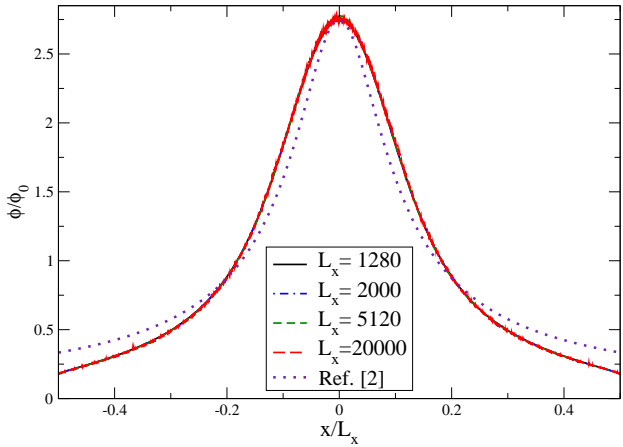}
    \caption{Master curve for rescaled local area fraction in
      different low-density systems  with the same line density
      [$\alpha = 0.9$, $\lambda =10.24$, $\ell /L_{x} \approx 0.03$].
      The dotted line corresponds to the theoretical 
      prediction of \cite{grossman97}.
      \label{fig:densityscaling}}
  \end{center}
\end{figure}

For low global area fractions, $\phi_{0} \lesssim 0.01$, we observe
scaling of the relative local area fraction $\phi(x)/\phi_{0}$ when
plotted versus $x/L_x$.  This is shown in
Fig.~\ref{fig:densityscaling}, where we compare density profiles of
systems with the same degree of inelasticity $\alpha=0.9$ and the same
line density $\lambda = 10.24$, but different values of the box width
$1280 \le L_x \le 20000$, corresponding to global area fraction
$\phi_{0}$ between $6\cdot 10^{-3}$ and $4\cdot 10^{-4}$ . The
relative local area fraction $\phi/\phi_{0}$ is seen to be a function
of $x/L_x$ only and does not depend separately on $L_x$ and $\phi_0$.
If the global area fraction were increased beyond $\phi_{0} \approx
0.01$ (not shown), then the data collapse would cease to hold. In addition,
the height of the peak would be reduced and instead of the bell-shaped 
master curve in Fig.~\ref{fig:densityscaling} one would get overall concave
profiles, similar to the one in Fig.~\ref{fig:dens.temp.kes.2}.

The hydrodynamic approach to quasi elastic, driven granular gases in
Sec.~II~D in \cite{grossman97} predicts for low-density systems that
the mean free path in units of $L_{x}$ depends only on the line
density according to $\ell/L_{x} = 1/(\sqrt{8}\lambda)$. When computed
for the situation of Fig.~\ref{fig:densityscaling}, one gets the small
numerical value $\ell/L_{x} \approx 0.03$, indicating that $\ell$ is
not a relevant length scale for the master curve of the rescaled
spatial density profiles. The theory of \cite{grossman97} also makes a
prediction for the relative local area fraction $\phi(x)/\phi_{0}$ in
terms of the solution of a first-order differential equation. This
differential equation includes one free parameter, which we have
fitted in order to match the solution (dotted line in
Fig.~\ref{fig:densityscaling}) at $x=0$ with the master curve from our
simulations. The agreement is reasonable, showing that inelasticities
of $\alpha=0.9$ are at the borderline of the scope of this otherwise
powerful approach to quasi elastic, driven granular gases.

\section{Local equation of state}
\label{sec:loc.equ.state}


\begin{figure}
  \begin{center}
    \psfrag{Enskog}{\tiny Eq.\ (\ref{eq:VEID}) with $\alpha$ = 0.9}
    \psfrag{Grossman}{\tiny Eq.\ (\ref{grossman}), elastic}
    \psfrag{Luding}{\tiny Eq.~(4) in \cite{luding01d}, elastic}
    \epsfig{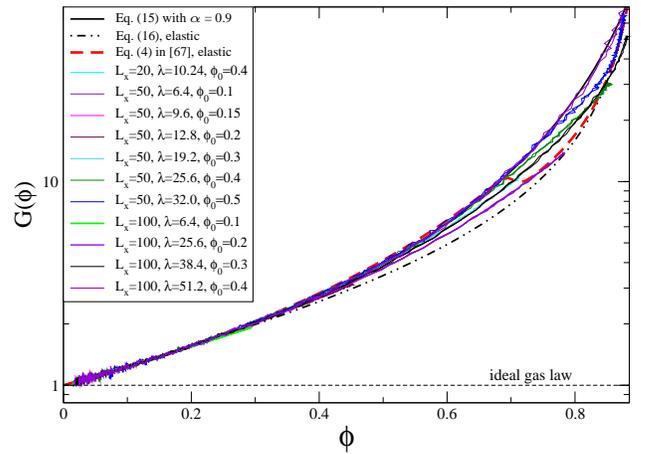} 
    \caption{(Color online) Semilogarithmic parametric plots of the function
      $G(\phi)$ from the local equation of state (\ref{eq:loceqnst})
      for different systems [$\alpha =0.9$, $L_{y}=25$, $0.01 \le
      \ell/L_{x} \le 0.05$]. Also shown are
      the theoretical predictions from (\ref{eq:VEID}),
      (\ref{grossman}), and Eq.~(4) in \cite{luding01d}.
      \label{fig:LocEquState}}
  \end{center}
\end{figure}

Having measured hydrodynamic fields in the steady state of the system, one is
led to search for a relation among them. Such a constitutive equation is
needed, for example, in all hydrodynamic approaches to driven granular gases
in order to obtain a closed set of equations \cite{grossman97,
  brey00c, livne02b}. In the last section 
it was shown that a driven granular gas is intrinsically inhomogeneous.
Therefore it is only natural to investigate how the \emph{local} values of the
granular temperature $T(x)$, pressure $p(x)$, and density $\rho(x)$---resp.\ 
area fraction $\phi(x) = \rho(x) \pi/4$ in our units---are related to each
other. To do so we observe in Fig.~\ref{fig:dens.temp.kes.2} that $T(x)$,
$p(x)$, and $\phi(x)$ are all symmetric in $x$. Moreover, and this is crucial, 
$\phi(x)$ is monotone in $x$ for either sign of $x$ (except for a boundary
layer of approximately one diameter in width close to a driving wall, which we
ignore). Therefore one can invert the function $\phi(x)$ for positive $x$.
Upon inserting $x = \phi^{-1}\bigl(\phi(x)\bigr)$ into the local pressure
and temperature, we arrive at the constitutive equation
\begin{equation}
  \label{eq:loceqnst}
  \frac{p(x)}{\rho(x) T(x)}= G\bigl(\phi(x)\bigr)
\end{equation}
with some function $G$.  Fig.~\ref{fig:LocEquState} shows parametric plots on
a semilogarithmic scale of the function $G$ with the values of
$p(x)/[\rho(x)T(x)]$ plotted against those of $\phi(x)$ for all $x$ (except
for those in the boundary layer mentioned above). In order to utilize a broad
range of $\phi$-values, Fig.~\ref{fig:LocEquState} contains data from 11
different systems, all of which have the same coefficient of restitution
$\alpha =0.9$ and the same height $L_{y}=25$, but different widths $L_{x}$ and
different global area fractions $\phi_{0}$. For not too large values of $\phi$
these data merge quite nicely, indicating that there is only a weak dependence
of $G$ on the global system parameters $L_{x}$ and $\phi_{0}$ in the
corresponding parameter range. In this case the
constitutive equation (\ref{eq:loceqnst}) can be interpreted as the
local equation of state of the 
system. The horizontal line in Fig.~\ref{fig:LocEquState} marks ideal-gas
behavior, from which $G$ deviates due to the collisional contribution to the
pressure.  These deviations increase significantly with increasing
$\phi$. 

The dotted line in Fig.~\ref{fig:LocEquState} corresponds to the function 
\begin{equation}
  \label{eq:VEID}
  G(\phi) = 1+ (1+\alpha) \phi \, \chi\,,
\end{equation}
where $\chi$ stands for the pair correlation function at contact of
the associated elastic hard-sphere gas in thermal equilibrium
\footnote{The expression (\ref{eq:VEID}) for $G$ with $\alpha =1$ is
  known from the exact equation of state $p/(\rho T) = 1+
  2^{d-1}\phi\, \chi$ of a $d$-dimensional elastic hard-sphere gas in
  thermal equilibrium \cite{hansen86}.}. Since $\chi$ is not known
exactly, we estimate it by the Henderson approximation
(\ref{chiapprox}) for numerical purposes. In the context of granular
gases, (\ref{eq:VEID}) occurred originally in a global equation of
state for the homogeneous situation of a freely cooling granular gas
\cite{goldshtein95}.  In \cite{soto01} the pair correlation function
has been studied in a homogeneously driven inelastic system with
periodic boundary conditions.  It was found to be nearly independent
of the coefficient of restitution and well approximated by the
Henderson approximation, Eq.  (\ref{chiapprox}). In \cite{barrat02b}
this form of $G$ was used in the local equation of state
(\ref{eq:loceqnst}) to get a theoretical prediction for $p(x)$ from
simulations of $T(x)$ and $\phi(x)$ in a driven granular gas.
Fig.~\ref{fig:LocEquState} reveals that this works generally quite
well for up to rather high local area fractions $\phi(x) \lesssim
0.5$. In even denser systems agreement still holds for the well
fluidized parts. Deviations from (\ref{eq:VEID}) start to occur when
entering the transition zones to the frozen-out stripe of particles in
the center of these high-density systems.  The dashed-dotted line in
Fig.~\ref{fig:LocEquState} corresponds to the interpolation formula
\cite{grossman97}
\begin{equation}
  \label{grossman}
  G(\phi) = \frac{\phi_{c} + \phi}{\phi_{c} - \phi}
\end{equation}
for $0 \le \phi < \phi_{c}$, which connects the behavior of dilute
(van der Waals) and dense (ordered) elastic hard-sphere systems. Here
$\phi_{c} := \pi/(2\sqrt{3}) \approx 0.91$ denotes the area fraction
for ordered closed packings in two dimensions. Eq.~(\ref{grossman})
was applied to quasi-elastic granular gases in \cite{grossman97}, and
even for our simulations with $\alpha = 0.9$ there is agreement with
the local data in the low-density regions up to $\phi(x) \lesssim
0.4$. In addition, the most dense regions of regularly ordered,
frozen-out particles in the center of the high-density systems are
described correctly, too.  Yet another interpolation formula for $G$,
which is rather accurate for an elastic hard-sphere gas even in the
vicinity of the freezing transition, was put forward in Eq.~(4) in
\cite{luding01d} and is depicted by the dashed line in
Fig.~\ref{fig:LocEquState}. Thus, the crossover between fluidized and
frozen-out behavior in the inelastic, driven systems, originating from
the transition zone at the border of the solidified stripe of
particles in the system center, is very smooth in comparison to the
coexistence region of the freezing transition for an equilibrated
elastic gas. Moreover, the location of the crossover depends on the
global system parameters.  Since both interpolation formulae,
(\ref{grossman}) and the one from \cite{luding01d}, were tailored for
elastic systems in thermal equilibrium, it is surprising to find such
a good agreement with the local data for driven inelastic hard-sphere
gases with restitution $\alpha =0.9$.

\begin{figure}
  \begin{center}
    \psfrag{Enskog}{\tiny Eq.\ (\ref{eq:VEID}) with $\alpha$ = 0.5}
    \psfrag{Grossman}{\tiny Eq.\ (\ref{grossman}), elastic}
    \epsfig{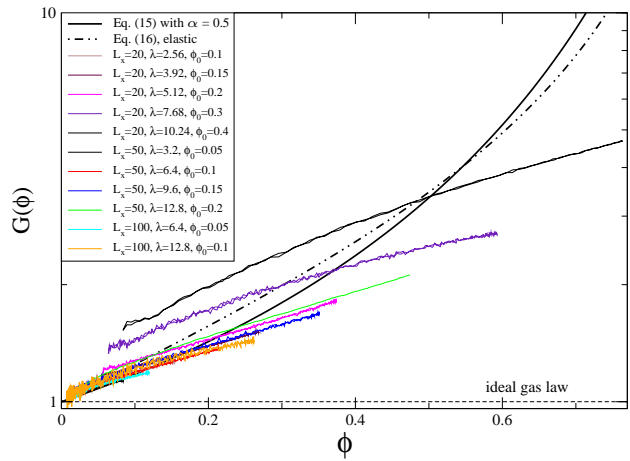} 
    \caption{(Color online) Same as Fig.~\ref{fig:LocEquState} but for
      more inelastic systems [$\alpha =0.5$, $L_{y}=25$, $0.01 \le
      \ell/L_{x} \le 0.1$].
      \label{fig:LocEquState-05}}
  \end{center}
\end{figure}

When lowering the coefficient of restitution  $\alpha$ from $0.9$ to
$0.8$ (not shown), there are only little changes to the plot in
Fig.~\ref{fig:LocEquState}. In particular, the spread of the data
pertaining to different systems increases, even for low
densities. Still one may
be inclined to interpret (\ref{eq:loceqnst}) as a local equation of
state---at least approximately. This situation is intermediate to the
previous with $\alpha =0.9$ and the following for strongly 
inelastic systems with coefficient of restitution  $\alpha =
0.5$. Indeed, Fig.~\ref{fig:LocEquState-05} reveals major discrepancies in the
local pressure, which are due to the systems with global
area fractions above $\phi_{0} \approx 0.2$. The discrepancies occur even at
positions in the sample where the local area fractions are below $0.2$ and
where all the more dilute systems, i.e.\ those with $\phi_{0}\le 0.2$,
agree reasonably with the proposal (\ref{eq:VEID}). The concept of a
local equation of state is therefore not sustainable any more for such strongly
inelastic systems, and (\ref{eq:loceqnst}) merely plays the role of a
local constitutive equation, which depends in addition on the global
system parameters.

\section{Absence of scaling for velocity distributions}
\label{sec:vel.distr.}


Finally, we examine the local velocity distributions of the driven
granular gas in the stationary state. We distinguish
between the local distribution
\begin{equation}
  \label{eq:fxdef}
  f_{x}(x,v_{x}) := \frac{1}{\rho(x)} \; \int_{\mathbb{R}}\! dv_{y} \;
  f_{\mathrm{stat}}(x,v_{x},v_{y}) 
\end{equation}
at position $x$ of the velocity component $v_{x}$ in the direction of
the driving and the distribution
\begin{equation}
  \label{eq:fydef}
  f_{y}(x,v_{y}) := \frac{1}{\rho(x)} \;\int_{\mathbb{R}}\! dv_{x} \;
  f_{\mathrm{stat}}(x,v_{x},v_{y}) 
\end{equation}
of the velocity component $v_{y}$ perpendicular to the direction of
the driving. By definition, these velocity distributions are normalized
to unity.
 
In order to determine $f_{x}$ and $f_{y}$ from the simulation we use
two different methods. The first one extracts them directly according
to their definitions (\ref{eq:fxdef}) and (\ref{eq:fydef}) from
$f_{\mathrm{stat}}$, which is determined as described in
Subsec.~\ref{sec:observables}. One disadvantage of this method is that
it yields a smeared-out velocity distribution, which is spatially
averaged over the width of the strip $\mathcal{V}_{x}$ centered around
$x$. This is of practically no importance in the middle of the
simulation box, but strongly disturbing for resolving the subtleties
which occur close to the driving walls and are presented in
Subsec.~\ref{sec:discontv}. The second way of measuring the velocity
distributions avoids this problem: To get $f_{i}$ ($i=x$ or $y$) we
keep track of all those particles which pass the line parallel to the
$y$-axis at position $x$ within a very long time interval of length
$\tau$ and whose $i$th component of the velocity lies in a
small interval of width $\Delta v$ around $v_{i}$.  Let us enumerate
these particles by $n_{i}$ and denote their velocity components by
$v_{x}^{(n_{i})}$ and $v_{y}^{(n_{i})}$.  Then one has (in the limits
$\tau\to\infty$ and $\Delta v\to 0$)
\begin{equation}
  \label{eq:method2}
  \rho(x) \,f_{i}(x,v_{i})  = \frac{1}{\tau \Delta v L_{y}}
  \sum_{n_{i}}  
  \frac{1}{|v_{x}^{(n_{i})}|} \,,
\end{equation}
because for $i=x$, resp.\ $i=y$, the right-hand side of
(\ref{eq:method2}) is equal to
\begin{equation}
  \label{eq:understandingmethod2}
  \int_{\mathbb{R}} \! dv_{y}\;
  \frac{|F_{x}(x,v_{x},v_{y})|}{|v_{x}|}
  \qquad
  \mbox{resp.~~}  \int_{\mathbb{R}} \! dv_{x}\;
    \frac{|F_{x}(x,v_{x},v_{y})|}{|v_{x}|} \,.
\end{equation}
Here, $F_{x}(x,v_{x},v_{y}) := v_{x} f_{\mathrm{stat}}(x,v_{x},v_{y})$
denotes the $x$-component of the (differential) current density in the
stationary state at position $x$ of particles with velocity components
$v_{x}$ and $v_{y}$.
 
As compared to the first method of measuring velocity distributions,
this one is also statistically more effective for determining rare
events, such as the high-velocity tails in
Subsec.~\ref{sec:novscaling}. However, as far as $f_{x}(x,v_{x})$ is
concerned for small $v_{x}$, the second method is inferior to the
first, because (\ref{eq:method2}) assigns a large weight to the
relevant events and therefore amplifies statistical fluctuations, too.


\subsection{Effects of the discontinuity at a driving wall}
\label{sec:discontv} 

 
Particle-number conservation at a driving wall requires the incoming particle
flux at the wall to be equal to the outgoing flux. For the velocity
distribution $f_{x}$ this implies the boundary condition \cite{brey00c}
\begin{align}
  \label{eq:RB} f_x(\mp {L_x}/{2},v_x) = & \Theta(\pm v_x - v_{\rm
    drive}) \, \biggl(1 \mp \frac{v_{\rm drive}}{v_x} \biggr) \nonumber\\
  & \times f_x(\mp {L_x}/{2},-v_x+v_{\rm drive})\,,
\end{align}
which must hold for all $v_{x}>0$. Here $\Theta(z):=1$ for $z \ge 0$,
respectively $\Theta(z):=0$ for $z<0$, denotes Heaviside's unit-step function
and $v_{\rm drive}=1$ in our units chosen. The boundary condition
(\ref{eq:RB}) relates the distribution $f_x$ of velocities prior to a
collision with the wall to the one after a collision with the wall.
In contrast, the velocity distribution $f_{y}$ must obey the usual
reflection symmetry in $v_{y}$ everywhere in the system, that is
\begin{equation}
  f_y(x,v_y) = f_y(x, - v_y) 
\end{equation}
for all $|x| \le L_{x}$ and all $v_{y} >0$.

\begin{figure}
  \begin{center}
    \leavevmode 
    \psfrag{fx*sqrt(Tx)}{ } 
    \psfrag{x/Lx}{$x/L_x$}
    \psfrag{vx/sqrt(Tx)}{$v_x/\sqrt{T_x(x)}$ }
    \psfig{file=\picdirectory/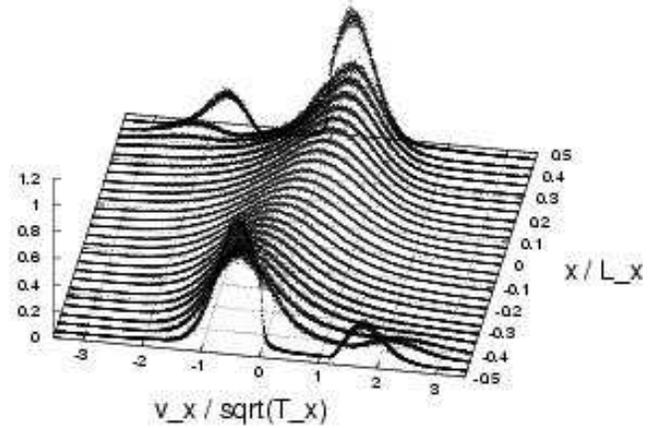, clip=, width=\columnwidth}
    \caption{Rescaled velocity
      distributions $\tilde{f}_x$ for different $x$.  [System
      parameters: $\alpha=0.9$, $L_{x}=20$, $\lambda = 10.24$
      ($N=256$, $\phi_0=0.4$)]
      \label{fig:Propagation_vx}}
  \end{center}
\end{figure}
 
\begin{figure}
  \begin{center}
    \leavevmode \psfrag{vx/sqrt(Tx)}{$v_x/\sqrt{T_x(x)}$}
    \epsfig{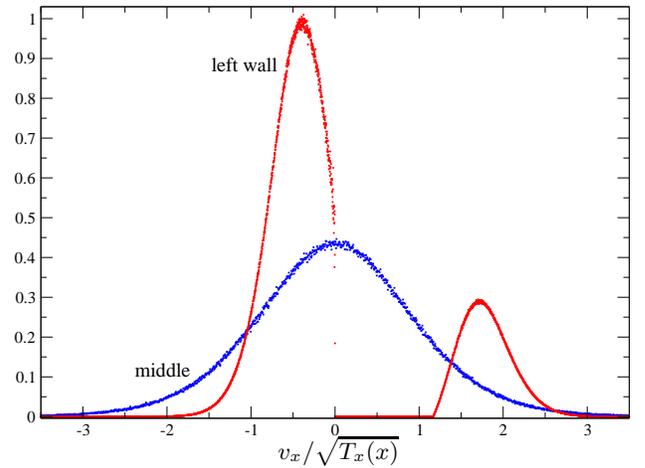}
    \caption{Two rescaled velocity distributions $\tilde{f}_x$ from
      Fig.~\ref{fig:Propagation_vx}: at the left wall and in the
      middle of the system.
      \label{fig:vx}}
  \end{center}
\end{figure}

We measured the velocity distribution $f_x(x, v_x)$ at 25 different
positions in a moderately inelastic system with coefficient of
restitution $\alpha =0.9$.  Fig.~\ref{fig:Propagation_vx} shows the
rescaled velocity distribution $\tilde{f}_x(x, v_x/\sqrt{T_x(x)}) :=
f_x(x,v_x) \sqrt{T_x(x)}$, measured using method one. At the driving
walls $\tilde{f}_x$ is seen to obey the boundary condition
(\ref{eq:RB}). When moving from a driving wall towards the center of
the system, the gap in $\tilde{f}_{x}$ gets gradually smeared out, and
$\tilde{f}_{x}$ becomes more and more symmetric.  Clearly, even for
this moderately inelastic system $\tilde{f}_x$ does not scale for
different $x$.  The two extreme cases for $x=-L_x/2$ and $x=0$ are
shown together in Fig.~\ref{fig:vx}. The data for $\tilde{f}_x$ at the
wall was obtained with the second method for measuring velocity
distributions. This result is in agreement with
Direct-Simulation-Monte-Carlo results in \cite{brey00c} and
Molecular-Dynamics simulations in \cite{barrat02b}.

\subsection{High-velocity tails} 
\label{sec:novscaling}


\begin{figure}
  \begin{center}
    \leavevmode 
    \psfrag{vi/sqrt(Ti)}{$v_i/\sqrt{T_i(x)}$}
    \psfrag{fx}{\footnotesize ${\tilde f}_x$}
    \psfrag{fy}{\footnotesize ${\tilde f}_y$} 
    \psfrag{fxx}{\tiny ${\tilde f}_x$} 
    \psfrag{fyy}{\tiny ${\tilde f}_y$}
    \epsfig{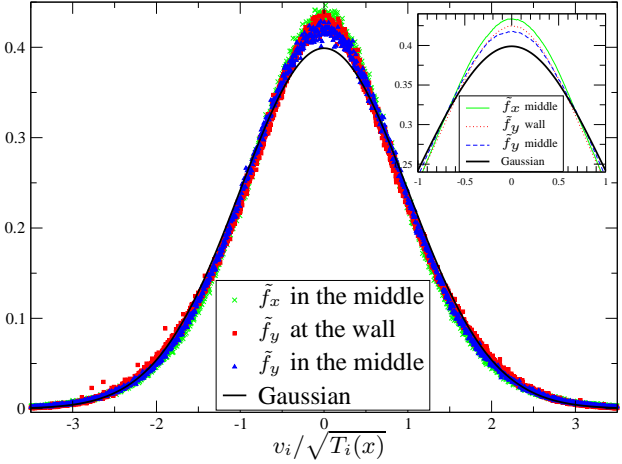}
    \caption{(Color online) Combined plot of rescaled velocity
      distributions of the same system as in
      Figs.~\ref{fig:Propagation_vx} and \ref{fig:vx}: $\tilde{f}_x$
      in the middle of the sample (green crosses), ${\tilde f}_y$ at a
      driving wall (red squares), and ${\tilde f}_y$ in the middle
      (blue triangles).  Also shown is a centered Gaussian with unit
      variance (black solid line).  The inset shows the central part
      of the main graph with the (1433/2964/1340) data points being
      smoothed (running average over (41/85/38) data points).
      \label{fig:vy.lin}}
  \end{center}
\end{figure}

\begin{figure*}
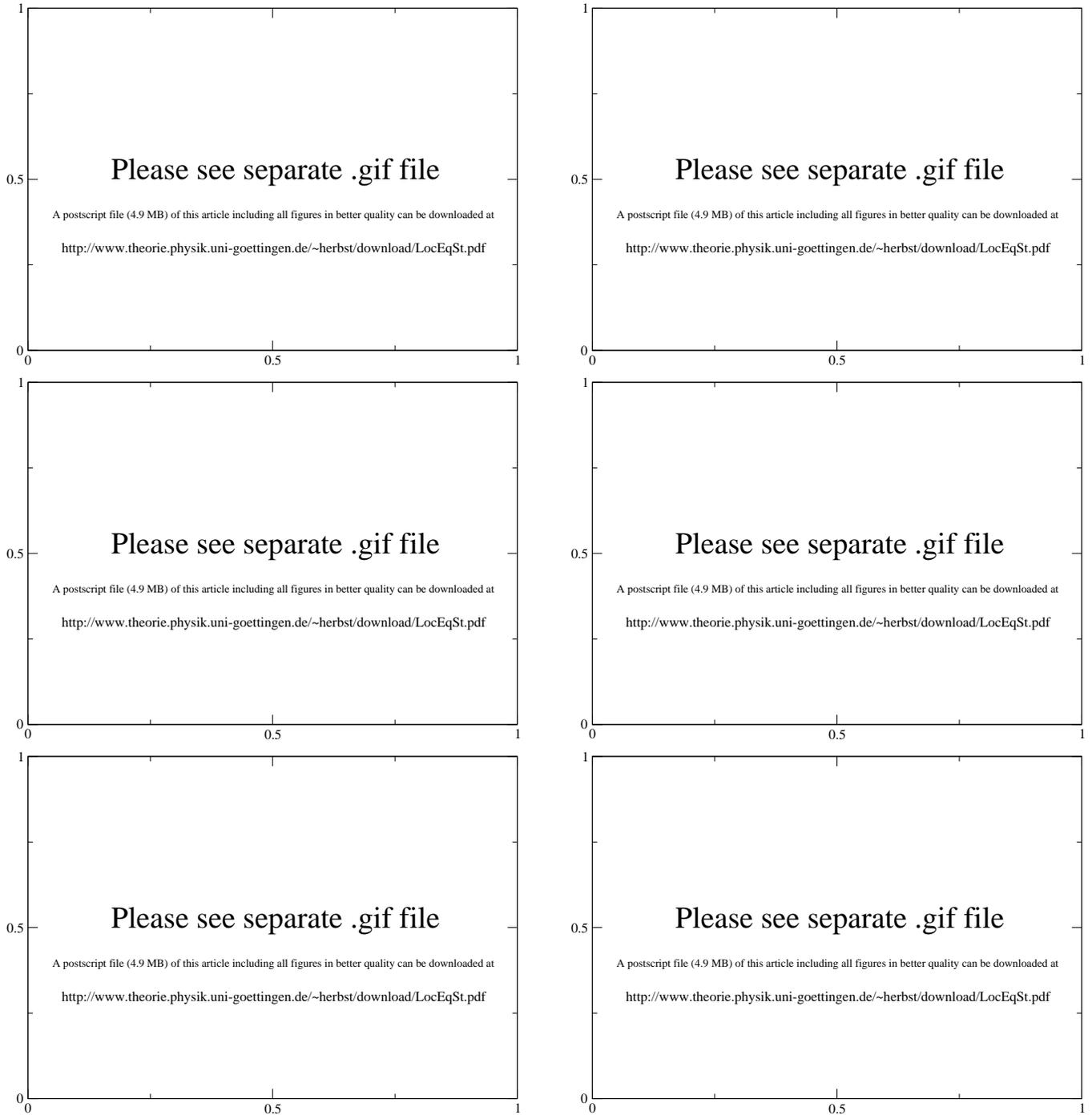

  \begin{center}
    \leavevmode 
    \psfrag{vi/sqrt(Ti)}{\footnotesize $v_i/\sqrt{T_i(x)}$}
    \psfrag{GAUSS}{\footnotesize Gauss}
     \epsfig{file=\picdirectory/empty.eps, clip=,
       width=.99\columnwidth}
     \hfill
     \epsfig{file=\picdirectory/empty.eps, clip=,
       width=.99\columnwidth} \\[1ex]
     \epsfig{file=\picdirectory/empty.eps, clip=,
       width=.99\columnwidth}
     \hfill
     \epsfig{file=\picdirectory/empty.eps, clip=,
       width=.99\columnwidth} \\[1ex]
     \epsfig{file=\picdirectory/empty.eps, clip=,
       width=.99\columnwidth}
     \hfill
      \epsfig{file=\picdirectory/empty.eps, clip=,
       width=.99\columnwidth}
    \caption{(Color online) Rescaled velocity distributions
      $\tilde{f}_x$ in the middle of the sample (green crosses),
      ${\tilde f}_y$ at a driving wall (red squares), and ${\tilde
        f}_y$ in the middle (blue triangles) as in
      Fig.~\ref{fig:vy.lin}, but here smoothed data are shown on a
      semi-logarithmic scale and velocities of much higher absolute
      values are included. The solid line is a centered Gaussian with
      unit variance. Different parts of the figure represent different
      systems with decreasing global area fraction $\phi_{0}$ (left to
      right) and decreasing coefficient of restitution $\alpha$ (top
      to bottom). The remaining system data are $L_{x}=20$, $\lambda =
      10.24$ ($N=256$) for the left column and $L_{x}=50$, $\lambda =
      9.6$ ($N=240$) for the right column.  The insets show
      $|\ln\tilde{f}_{i}|^{-1}$ on a double-logarithmic scale to
      determine the decay exponent $\beta$ from (\ref{betadef}). The
      dashed line is the best linear fit to $\tilde{f}_{x}$ in the
      middle of the sample, the solid line to $\tilde{f}_{y}$ in the
      middle, and the dotted line to $\tilde{f}_{y}$ at the wall.
      Part~a) corresponds to the system of Fig.~\ref{fig:vy.lin}.
      \label{fig:vy.log} 
    }
  \end{center}
\end{figure*}
 
Our main result for the velocity distributions is the non-scaling and
multiformity of their tails. In order to observe these phenomena,
extensive simulations for capturing rare events are required and the
data have to be analyzed on a logarithmic scale. In contrast, on a
\emph{linear} scale the rescaled velocity distributions seem to
collapse approximately, as was observed previously by e.g.\ 
\cite{grossman97, moon03}. As an example we show in
Fig.~\ref{fig:vy.lin} the rescaled distributions
$\tilde{f}_{y}(x,v_{y}/\sqrt{T_y(x)}) := f_y(x,v_y)\sqrt{T_y(x)}$ in
the middle of the sample ($x=0$) and at a driving wall ($x=\pm
L_{x}/2$) for the same moderately inelastic system ($\alpha = 0.9$) as
in Figs.~\ref{fig:Propagation_vx} and~\ref{fig:vx}.  Clearly, the
curves are reflection-symmetric in $v_{y}$.  For comparison, we have
also included the symmetric distribution $\tilde{f}_{x}$ in the middle
of the system. Deviations are visible for small velocities only, as
the inset of Fig.~\ref{fig:vy.lin} shows. The corresponding Gaussian
(solid line) in Fig.~\ref{fig:vy.lin} suggests that the velocity
distributions are close to but not identical to a Maxwellian. The
approximate data collapse, observed on this level of accuracy in
Fig.~\ref{fig:vy.lin}, even continues to hold if the coefficient of
restitution of the gas is varied in a not too large extent. In
considerably more inelastic systems, such as for $\alpha= 0.5$, this
is not true any longer. For example, the peak of $\tilde{f}_{x}$
measured in the middle of the system would be considerably broader and
flatter than the ones of $\tilde{f}_{y}$ in the center and at the wall
(not shown).
 
\begin{figure}
  \begin{center}
    \leavevmode 
    \psfrag{vi}{$v_i$}
    \psfrag{fxmiddle}{\raisebox{.4cm}{\footnotesize
        \hspace*{-0.4cm}$\sqrt{T_{x}(0)} f_{x}(0, v_{x})$}} 
    \psfrag{fymiddle}{\raisebox{.1cm}{\footnotesize $\sqrt{\smash[b]{T_{y}(0)}}
      f_{y}(0, v_{y})$}} 
    \psfrag{fywall}{\raisebox{.4cm}{\footnotesize \hspace*{1.5cm}
      $\sqrt{\smash[b]{T_{y}(L_{x}/2)}} f_{y}(L_{x}/2, v_{y})$}} 
    \epsfig{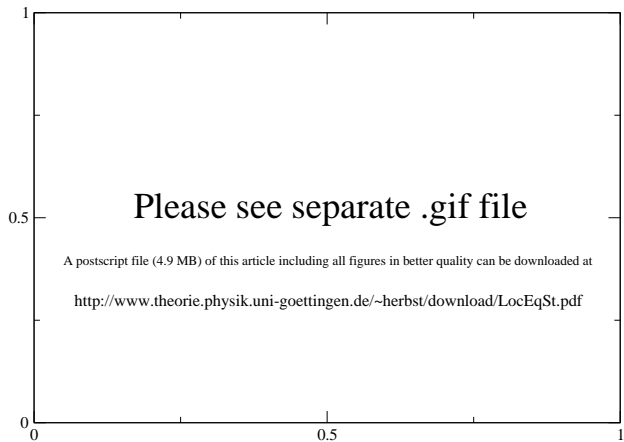}
    \caption{(Color online) Velocity distributions for the system in 
      Fig.~\ref{fig:vy.log}~f), but with unsmoothed data and without
      rescaling of the horizontal axis.
      \label{fig:vy.original}}
  \end{center}
\end{figure}

In contrast, Fig.~\ref{fig:vy.log}~a) shows the same data of
Fig.~\ref{fig:vy.lin} on a \emph{semi-logarithmic} scale and includes
also velocities of much higher absolute values. From this figure it is
evident that scaling does not hold in the high-velocity tails of the
distributions either. Similar observations were made before in e.g.\ 
\cite{barrat02b, zon03, brey03a}. The type of decay in the
high-velocity tails is different for $\tilde{f}_x$ and $\tilde{f}_y$,
and also depends on the position in the sample, the coefficient of
restitution $\alpha$, and the global area fraction $\phi_{0}$. This is
illustrated by examples of different systems in Fig.~\ref{fig:vy.log}.
In the insets we show $|\ln \tilde{f}_{i}|^{-1}$ on a
double-logarithmic scale in order to determine the decay exponent
$\beta$ defined by
\begin{equation}
  \label{betadef}
  \ln \tilde{f}_{y}\bigl(0, v_{y}/ \sqrt{\smash[b]{T_{y}(0)}}\bigr)
   \stackrel{|v_{y}| \to \infty}{\sim}\;  - 
  \bigl|v_{y}/ \sqrt{\smash[b]{T_{y}(0)}} 
  \bigr|^{\beta} 
\end{equation}
for $\tilde{f}_{y}$ in the middle of the sample. The exponent is
defined accordingly for $\tilde{f}_{y}$ at the wall and
$\tilde{f}_{x}$ in the middle of the sample. For the moderately
inelastic systems with $\alpha = 0.9$ in the first row of
Fig.~\ref{fig:vy.log}, the asymptotics has been clearly reached. We
note that $\beta$ is different for the different distributions, and
also depends on the global area fraction $\phi_{0}$.  Upon lowering
$\alpha$ (top to bottom in Fig.~\ref{fig:vy.log}) and/or decreasing
the global area fraction $\phi_{0}$ (left to right), the tails of
$\tilde{f}_{y}$ get more and more populated, that is, $\beta$
decreases.  In very simple terms, this may be understood from the fact
that (i)~the largest typical velocities are always of the order of
$v_{\mathrm{drive}}=1$ (see also Fig.~\ref{fig:vy.original}) and
(ii)~that $v_{\mathrm{drive}}/\sqrt{T_{i}(x)}$ increases up to 20 with
decreasing $\alpha$ and decreasing $\phi_{0}$. Hence, a Maxwellian
velocity distribution would not be able to supply enough probability
to particles with velocities of the order of $v_{\mathrm{drive}}$,
instead higher-populated tails are needed. This argument suggests
different behavior in different velocity regions so that the
distributions cannot be fitted to the functional form (\ref{betadef})
over the entire range of velocities \cite{zon03}.  Indeed, such a
behavior can be seen in Fig.~\ref{fig:vy.log}~d), e) and~f). The final
asymptotics could not always be deduced from the simulations, even
though our data include velocities which are up to 40 times bigger
than the appropriate granular velocities $\sqrt{T_{i}}$.  This applies
to $\tilde{f}_{y}$ in the middle of the sample in part~e), where we
suspect that the final asymptotics has not been reached. An asymptotic
analysis of $\tilde{f}_{x}$ in the middle of the sample is even more
problematic due to particles that reach the system center from a
driving wall without undergoing a collision. These particles give rise
to the side peaks of $\tilde{f}_{x}$ in parts~e) and~f), which have
prevented us from determining $\beta$ in these cases.

Fig.~\ref{fig:vy.log} contains smoothed data as in the inset of
Fig.~\ref{fig:vy.lin}, and the scales of the horizontal axes are
determined by the square root of the appropriate granular
temperatures. For comparison, Fig.~\ref{fig:vy.original} shows the
unsmoothed data corresponding to Fig.~\ref{fig:vy.log}~f), plotted
directly versus $v_{i}$ (in units of $v_{\mathrm{drive}}$). The side
peaks of $\tilde{f}_{x}$ in the middle of the sample are due to the
above-mentioned particles which fly from a driving wall to the system
center without undergoing a collision. For the vast majority of
particles $|v_{y}|$ is less than 80\% of the driving velocity. It is
in the region of this highest velocity observed for $\tilde{f}_{y}$ in
the middle of the sample, where $\tilde{f}_{x}$ changes abruptly its
slope.

The exponent $\beta$ has also been determined experimentally in a
strongly driven gas so that gravity effects are small \cite{rouyer00}.
A value $\beta \approx 1.55$ was measured for a gas with coefficient
of restitution $\alpha \approx 0.93$. It was found to be remarkably
independent of the global area fraction $\phi_{0}$, which was varied
from $0.05$ to $0.3$. We have also simulated the system of
\cite{rouyer00} with zero gravity (not shown) and
reproduced their value for $\beta$. Even though the experimental data
cover a wide range of velocities, some of the interesting phenomena
discussed in Fig.~\ref{fig:vy.log} cannot be observed in this range.
For the same reason the significance of the particular value $\beta
\approx 1.55$ should not be overestimated.

Simulations of a driven granular gas in a circularly shaped box also
yield stretched Gaussian tails \cite{zon03}. In addition, evidence is
given that $\beta$ depends only on the coefficient of restitution and
the average ratio of the number of particle-wall collisions and
particle-particle collisions.  For homogeneously driven systems the
tail of the velocity distribution was theoretically predicted
\cite{noije98d} to be governed by the decay exponent $\beta =3/2$.
Simulations of homogeneously driven systems \cite{moon01,barrat03}
observed the exponent $\beta =3/2$ only for unrealistically low values
of the coefficient of restitution.


\section{Summary and Outlook}
\label{sec:conclusions}


In this article we explored the steady-state properties of a granular
gas driven by vibrating walls. We have measured the full stress
tensor, including the collisional contribution. This allowed us to
obtain a constitutive equation directly from the simulations. The
constitutive equation relates local density, pressure and temperature.
For small inelasticities it can be regarded as the local equation of
state of the gas, because it is to a large extent independent of the
global system parameters. For strongly inelastic systems this
interpretation cannot be sustained, instead the constitutive relation
depends on the global volume fraction and sample geometry.  We have
also measured local velocity distributions, whose high-velocity tails
were found to depend on the position in the sample as well as on the
coefficient of restitution and the global volume fraction. Moreover,
the tails are different for the two velocity components parallel and
perpendicular to the driving walls.  To conclude, the stationary state
of a driven granular gas is in general {\it non-universal}---in
contrast to the corresponding elastic system.  This is not unexpected
because the driven granular gas is a non-equilibrium state.
Furthermore driving the system by energy input through the walls is
effective only if the distance between the driving walls is finite so
that the sample geometry enters naturally.

We plan to extend our studies in various directions. It is
straightforward to also consider rotational motion of the disks, by
generalizing the collision rules to include tangential restitution as
well as Coulomb friction. Again, the question arises to what extent
the additional parameters affect the local equation of state.  In this
context it would be interesting to study the effects of gravity for
better comparison with experiments.  Another interesting quantity to
measure is the heat flux which, like the stress tensor, is also
expected to be affected by collisions.  Finally, one might also
investigate other driving mechanisms, like vibrations with finite
frequency and non-zero amplitude or sinusoidal driving. Work along
these lines is in progress.

\begin{acknowledgments}
  We are grateful to Timo Aspelmeier for providing us with the first version
  of our program. Olaf Herbst thanks Stefan Luding and Isaac Goldhirsch for
  inspiring discussions. Finally, we acknowledge financial support by the DFG
  through SFB~602 and Grant No.\ Zi~209/6--1.
\end{acknowledgments}


\appendix*
\section{A simple energy-balance argument}


Here we derive the energy balance Eq.\ (\ref{eq:EnBilanzGlSimple}) by
using arguments as in \cite{kumaran98, mcnamara98b}.
We do this in a slightly more general setting than needed for
Eq.\ (\ref{eq:EnBilanzGlSimple}) and allow in addition for inelastic
collisions with the wall, characterized by a coefficient of restitution
$\alpha_{\rm w}$. The appropriate generalization of the collision rule
(\ref{eq:freddyandDeltasWall}) includes both the driving velocity and
the coefficient of restitution with the wall,
\begin{equation} 
  \label{general.wall.coll} 
  \bm{v}' = \bm{v}+ \Delta \bm{v}_{\rm pw} \quad 
  \mbox{with} \quad \Delta \bm{v}_{\rm pw} =
  [- (1+ \alpha_{\rm w}) v_x \pm v_{\rm drive}] \bm{e}_{x}\,.
\end{equation}
The special case $v_{\rm drive}=0$ and $\alpha_{\rm w}>1$ provides an
alternative driving mechanism, which, however, does not give rise to a
stationary state, as will be shown below. In order to be able to treat
this limit and for a better readability of
the formulae, we refrain from using dimensionless physical units in
this appendix.

The average energy gain $\Delta E_{\rm pw}$ due to a particle-wall
collision is estimated from Eq.\  (\ref{general.wall.coll}) by
averaging the kinetic energy before and after the collision with a
Maxwellian velocity distribution with (global) temperature $T$. This
gives
\begin{equation}
  \label{eq:appendix.DeltaE}
  \Delta E_{\rm pw} = \frac{m}{2}
  \biggl( v_{\rm drive}^2+ 4 \alpha_{\rm w} \sqrt{\frac{T}{2\pi m}}
    v_{\rm drive} - (1- \alpha^2_{\rm w}) \frac{T}{m} \biggr)\,.
\end{equation}
The collision frequency of particles with the left (right) wall is estimated by
\begin{equation}
  f_{\rm pw} =  \frac{N}{L_{x}}\, \sqrt{\frac{T}{2\pi m}}\,,
\end{equation}
where we have assumed the density to be spatially homogeneous
throughout the system.  

When two disks collide in the bulk, the average change in total energy
is computed similarly to (\ref{eq:appendix.DeltaE}) from
(\ref{eq:freddy}) and (\ref{eq:Deltas}) by 
averaging pre- and post-collisional kinetic energy, which yields
\begin{equation}
  \Delta E_{\rm pp} =  -  \frac{1-\alpha^2}{2}\; T\,.
\end{equation}
Finally, the number of particle-particle collisions per time is given
approximately by Enskog's collision frequency \cite{noije98d}
\begin{equation}
  f_{\rm pp} = \frac{N}{L_{x}}\, \lambda a \chi \sqrt{\frac{T\pi}{m}} \,,
\end{equation}
where $\chi$ is the pair correlation function at contact of a
corresponding elastic gas in thermal equilibrium. Since $\chi$ is not
exactly known, we resort to the widely used Henderson approximation
\cite{henderson75}
\begin{equation}
  \label{chiapprox}
  \chi \approx \frac{1 - 7\phi/16}{(1-\phi)^{2}}
\end{equation}
for numerical purposes. It may be viewed as a heuristic rational
approximation to the virial expansion of $\chi$ and is the
two-dimensional equivalent to the Carnahan-Starling approximation
\cite{hansen86} for a three-dimensional hard-sphere gas. Additional
higher-order terms to the Henderson approximation, which are
proportional to $\phi^{3}/(1-\phi)^{4}$, have turned out to be
irrelevant for our purposes and will therefore not be taken into
account.

Summing over energy loss in the bulk and energy gain at the right and
left wall, we obtain for the total change in granular temperature
\begin{align}
  \label{eq:appendix.EnergieBilanzRoh}
  \frac{dT_{\rm}}{dt} &\approx  \frac{2f_{\rm pw}E_{\rm pw}+f_{\rm
      pp}E_{\rm pp}}{N} \nonumber\\
  &= \frac{m}{L_{x}}  \sqrt{\frac{T}{2\pi m}}
  \biggl( v_{\rm drive}^2  + 4 {\alpha}_{\rm w} \sqrt{\frac{T}{2\pi m}}
    v_{\rm drive}  - \frac{\pi \psi_{\mathrm{eff}} T}{2m}  \biggr)\,.
\end{align}
Here the dimensionless parameter
\begin{equation}
  \psi_{\mathrm{eff}} := \psi - \frac{2}{\pi}\, (\alpha_{\mathrm{w}}^{2}-1)
\end{equation}
is given in terms of 
$\psi := \sqrt{2} \lambda a\chi (1-\alpha^{2})$, which was already
introduced below Eq.~(\ref{eq:EnBilanzGlSimple}). (We note that
$\lambda a$ is the dimensionless line density employed there.)
We briefly discuss two special cases:

a) For $v_{\rm drive}=0$ and $\alpha_{\rm w}>1$ no
stationary state is reached in general. Both, energy gain and
loss increase like $T^{3/2}$, resulting in Haff's law
\begin{equation}
  \label{eq:appendix.essentiallyHaff}
  \frac{dT}{dt}  = - \frac{\psi_{\mathrm{eff}}}{2L_{x}} 
  \sqrt{\frac{\pi}{2m}} \, T^{3/2}\,,
\end{equation}
which has been discussed extensively in the different context of
freely cooling granular gases. Here the temperature continues to
decrease or increase depending on whether dissipation or driving wins.

b) For $\alpha_{\rm w}=1$ and $v_{\rm drive}>0$ the granular
temperature adjusts to the driving so that the stationary state with
$dT/dt =0$ is characterized by the quadratic equation
\begin{equation}
  \label{quadrat}
   1 + 2 \sqrt{\frac{2\overline{T}}{\pi}}
  - \frac{\pi\psi}{2} \, \overline{T} =0\,.
\end{equation}
The solution of (\ref{quadrat}) for the dimensionless global temperature
$\overline{T}:=T/(m v^2_{\rm drive})$ is given in
Eq.~(\ref{eq:EnBilanzGlSimple}).

\bibliographystyle{apsrev} 

\end{document}